\documentclass[a4paper,12pt]{article}
\usepackage{a4wide,latexsym,graphicx,epsfig,psfrag,float}
\usepackage{mathrsfs}
\usepackage{latexsym}
\usepackage{indentfirst}
\usepackage{amsmath}
\usepackage{amsxtra}
\usepackage{amsbsy}
\usepackage{amscd}
\usepackage{color}
\usepackage{hyperref}
\usepackage{multirow}

\usepackage{graphics}
\usepackage{amssymb}
\usepackage{amsfonts}

\usepackage{subfigure}
\usepackage{color}
\usepackage{times,fancyhdr}
\usepackage{amsfonts}

\allowdisplaybreaks

\newcommand{\rbox}{\rule[-0.20cm]{0cm}{8mm}}
\newcommand{\be}{\begin{equation}}
\newcommand{\ee}{\end{equation}}

\newcommand{\bea}{\begin{eqnarray}}
\newcommand{\eea}{\end{eqnarray}}
\newcommand{\nn}{\nonumber}

\def\no{\nonumber}
\def\bea{\arraycolsep .1em \begin{eqnarray}}
\def\eea{\end{eqnarray}}

\begin{document}

\begin{titlepage}
%\vskip 1cm
\begin{flushright}
\end{flushright}
\vspace{2cm}

\setcounter{footnote}{0}
\renewcommand{\thefootnote}{\fnsymbol{footnote}}

\begin{center}
{\LARGE \bf Pascalutsa-Vanderhaeghen light-by-light sumrule from photon-photon collisions}
\vspace{2cm} \\
{\sc  Ling-Yun Dai}$^{1,2,3}$\footnote{Email:~l.dai@fz-juelich.de}
{\sc , M.R. Pennington}$^{4,5}$\footnote{Email:~michaelp@jlab.org}
\vspace{0.8cm} \\
$^{1}$ {\small Institute for Advanced Simulation, Institut f\"ur Kernphysik
   and J\"ulich Center for Hadron Physics, Forschungszentrum J\"ulich, D-52425 J\"ulich,
 Germany} \\[1mm]
$^{2}$ {\small Center For Exploration of Energy and Matter, Indiana University, Bloomington, IN 47408, USA} \\[1mm]
$^{3}$ {\small Physics Department, Indiana University, Bloomington, IN 47405, USA}\\[1mm]
$^{4}$ {\small Thomas Jefferson National Accelerator Facility, Newport News, VA 23606, USA} \\[1mm]
$^{5}$ {\small Physics Department, College of William \& Mary, Williamsburg, VA 23187, USA}\\[1mm]
\end{center}

\setcounter{footnote}{0}
\renewcommand{\thefootnote}{\arabic{footnote}}
%\vspace{0.5cm}

\begin{abstract}
Light-by-light scattering sumrules based on general field theory principles relate cross-sections with different helicities.
In this paper the simplest sumrule is tested for the $I=0$ and  $2$ channels for \lq\lq real'' photon-photon collisions. Important contributions  come from the long-lived pseudoscalar mesons and
 from di-meson
intermediate states. The latest Amplitude Analysis of $\gamma\gamma\to\pi\pi, \overline{K}K$ allows this contribution to be evaluated.
However, we find that other multi-meson contributions up to 2.5~GeV are required to satisfy the sumrules. While data on three and four pion cross-sections exist,
there is no information about their isospin and helicity decomposition. Nevertheless, we show the measured cross-sections are sufficiently large to ensure
the sumrules for the helicity differences are likely fulfilled.
\end{abstract}
\vspace{0.5cm}

{\indent PACS~: 11.55.Fv, 14.40.Be, 11.80.Et \\
\indent Keywords~: Dispersion relations, Light mesons, Partial-wave analysis.}

\end{titlepage}

\section{Introduction}\label{sec:1}
There is keen interest in improving our understanding of light-by-light  scattering as an essential
ingredient of calculations of hadronic contributions to the anomalous magnetic moment of the muon in
preparation for planned experiments at Fermilab~\cite{Fermilab-g2} and J-PARC~\cite{J-PARC}. An essential component of this are tests of
the theoretical framework by the scattering of essentially real photons, as an anchor for  modeling
scattering with photons of virtuality up to 2~GeV$^2$ that control the multi-loop structure of hadronic light-by-light scattering.
Models of $\gamma^* \gamma^*$ scattering in different polarization states are expected to be constrained by
sumruies deduced by Pascalutsa and Vanderhaeghen (PV)~\cite{PV1} from general field theoretic considerations.
The $\gamma^*\gamma$ sumrules have been tested with new Belle data~\cite{Belle2016} and recently used to calculate the hadronic contribution to muon's anomalous magnetic
moment~\cite{Igor2016}.

Here we discuss what we currently know from the detailed analysis of all available data on two real photon interactions about the simplest of these sumrules.
For physical photons, the PV sumrules relate integrals of the total polarized and unpolarized cross sections to the low-energy structure of light-by-light scattering.
The simplest states that the helicity-two and helicity-zero cross-sections contribute
equally~\cite{PV1} so that the weighted integral from threshold $s_{th}$:
\be\label{eq:PV1}
\int_{s_{th}}^{\infty}\,ds\,\frac{\sigma_2\ -\sigma_0}{s}\;=\;0 \; ,
\ee
where the subscripts label the total helicity ($\lambda$) of the colliding photons. 
Subsequently, we denote the  difference  $[\sigma_2(s)\,-\,\sigma_0(s)]$ by $\Delta\sigma(s)$.
 This sumrule should be true for the sum of all hadronic intermediate states of definite isospin, {\it i.e} $I=0,\ 1$, and 2. Thus,
the first contributions to include in Eq.~(\ref{eq:PV1}) are from single particle intermediate states that appear  in $\gamma\gamma\to\gamma\gamma$ scattering, namely
the $\pi^0$ in the $I=1$ channel, and $\eta, \eta'$ in $I=0$. Their contribution to the helicity-zero cross-section is well-known
and included in Table~1, with uncertainties given by the decay rates from the PDG Review of Particle Properties~\cite{pdg}.

All the remaining contributions come from intermediate states that are multi-hadron channels, {\it e.g} $\pi\pi$, $3\pi$, $4\pi$ and so on, with kaons and protons replacing pions as the energy increases.
Some of these cross-sections have significant resonant contributions, for instance the neutral tensor mesons with the $f_2(1270)$ dominating the  $\pi\pi$ channel, the $a_2(1320)$ in  $3\pi$ and the $f_2^\prime(1525)$ in the ${\overline K}K$ channel. Their contributions  have been estimated in Ref.~\cite{PV2}, in the approximation that these resonances are narrow and only couple to photons with helicity-two. While these may seem plausible \lq\lq guesses'', it turns out in fact that they provide a rather poor description of the contribution of these spin two intermediate states.

 This fact highlights why
using published data on two or more particle production, it is not possible directly to evaluate the sumrule of Eq.~(\ref{eq:PV1}).
\begin{itemize}
\item First the observed cross-sections cover
only part of the angular range of the final state particles. When these are just two particles, this is typically limited to $|\cos \theta^*| \le 0.6$ for charged particles
and $\le 0.8$ for neutral, while the sumrule requires cross-sections integrated over the full angular range (where $\theta^*$ is the scattering angle in the $\gamma\gamma$ center-of-mass frame).
\item Secondly, measurements of all possible charged states are required to separate out the isospin components, even when there are just two final state particles.
\item Thirdly, in untagged electron-positron collisions there is no polarization information about the colliding
photons, which would automatically separate the helicity components in the sumrule of Eq.~(\ref{eq:PV1}).
\end{itemize}
Consequently, one needs to combine data with other information. The Amplitude Analyses performed in Refs.~\cite{mp1,boglione,lyd-mrp} address these issues by making use of the underlying $S$-Matrix principles of analyticity, crossing and unitarity, combined with the QED low energy theorem on Compton scattering. Without this technology,
a partial wave separation would not be possible. Even then this is limited to the c.m. energy region below ~1.44 GeV, beyond which multi-pion channels become crucially important: channels for which we have even more limited information from experiment, as we discuss in more detail below. We give the single and two particle contributions to the PV sumrule, Eq.~(1), that can be accurately computed in Section~2. Then in Section~3 we estimate the contribution of multiparticle and in Section~4 give our conclusions. 

\section{Contributions to the Sumrules: single particle and $\pi\pi$, ${\overline K}K$}

%It is helpful to note that as the cross-sections decrease modulo logarithms of $s$, like $1/s$ and the measure in Eqs.~(\ref{eq:PV1}-\ref{eq:PV2}) also has a factor of $1/s$. In addition there is a phase-space factor, which for two pion production is $\sqrt{1-4m_\pi^{\,2}/s}$. Consequently the integrand in both $\Sigma$ and $\Delta$ peak at an energy of $\sqrt{5} m_\pi$, {\it i.e.} 312~MeV.  This emphasises the importance of the low energy region for these integrals.

%The $\gamma\gamma\to\gamma\gamma$ cross-section in each helicity is related by the optical theorem to the imaginary part of the corresponding forward amplitude. The single particle contributions (of near stable) hadrons are readily computed. For the $I=0$ channel these are the $\eta$ and $\eta^\prime$, and for $I=1$ the $\pi^0$.  Their contribution is

We begin by considering the contributions to the PV sumrule, first from single particles in the process $\gamma\gamma\to\gamma\gamma$. By the optical theorem, the cross-section is related to the imaginary part of the relevant forward helicity amplitudes. Thus the PV sumrule involves the forward $f^{(-)}(s)$ amplitude, defined in ~\cite{PV1}, to be proportional to the difference of the ${\cal M}_{++++}-{\cal M}_{+-+-}$ amplitudes. Then the contribution of a near stable single particle of mass $M$ and $\gamma\gamma$ width $\Gamma_{\gamma\gamma}^{(\lambda)}$ in the helicity $\lambda$ channel is given by
\be
\sigma_\lambda(\gamma\gamma\to\gamma\gamma;s)\;=\;16\pi^2\,(2J+1)\frac{\Gamma_{\gamma\gamma}^{(\lambda)}}{M}\, \delta(s-M^2) \quad .
\ee
The contribution to the PV sumrule for the $\pi^0, \eta, \eta^\prime$ are readily deduced using this equation with the information from the PDG Tables~\cite{pdg}. These are listed in Table~1.
\begin{table}[ph]
\vspace{-0.0cm}
{\footnotesize
\begin{center}
\begin{tabular}  {|l|c|c|c|}
\hline
 contribution to $\Delta^{I}(4m_\pi^{\,2},2{\rm GeV}^2, Z=1)$& $I=0$   &  ${I=1}$   &  ${I=2}$   \rbox \\[0.5mm] \hline
$\gamma\gamma\to \pi^0$ ~\cite{pdg}~(nb) & -  &  -190.9$\pm$4.0                &  -                 \rbox \\[0.5mm] \hline
$\gamma\gamma\to \eta, \eta'$ ~\cite{pdg}~(nb) & -497.7$\pm$19.3  &  -                 &  -                 \rbox \\[0.5mm] \hline
$\gamma\gamma\to a_2(1320)$ ~\cite{pdg}~(nb) & -  &  {\it 135.0$\pm$12$\pm$25}~$\dag$               &  -                 \rbox \\[0.5mm] \hline
$\gamma\gamma\to\pi\pi$~(nb)          &     231.3$\pm$31.2      &  -                 &  -82.9$\pm$12.2     \rbox \\[0.5mm] \hline
$\gamma\gamma\to \overline{K}K $~(nb) &     6.2$\pm$2.0        &  0.9$\pm$0.2       &  -                 \rbox \\[0.5mm] \hline
                          SUM~(nb)    &    -260.2$\pm$36.7          &  -55.0$\pm$28.0                 &  -82.9$\pm$12.2     \rbox \\[0.5mm] \hline
\hline
evaluation of $\Delta^{I}(4m_\pi^{\,2},\infty, Z=1)$& $I=0$   &  ${I=1}$   &  ${I=2}$   \rbox \\[0.5mm] \hline
$\gamma\gamma\to \pi^0$ ~\cite{pdg}~(nb) & -  &  -190.9$\pm$4.0                &  -                 \rbox \\[0.5mm] \hline
$\gamma\gamma\to \eta, \eta'$ ~\cite{pdg}~(nb) & -497.7$\pm$19.3  &  -                 &  -                 \rbox \\[0.5mm] \hline
$\gamma\gamma\to a_2(1320)$ ~\cite{pdg}~(nb) & -  & {\it 135.0$\pm$12$\pm$25}~$\dag$               &  -                 \rbox \\[0.5mm] \hline
$\gamma\gamma\to\pi\pi$~(nb)          &     308.0$\pm$41.5      &  -                 &  -44.2$\pm$6.1     \rbox \\[0.5mm] \hline
$\gamma\gamma\to \overline{K}K $~(nb) &     23.7$\pm$7.5        &  18.1$\pm$4.9       &  -                 \rbox \\[0.5mm] \hline
                          SUM~(nb)    &    -166.0$\pm$46.4          & -37.8$\pm$28.4                 &  -44.2$\pm$6.1     \rbox \\[0.5mm] \hline
\end{tabular}
\caption{\label{tab:PV}PV sumrule contributions for intermedate states $\eta$, $\eta'$, $\pi\pi$ and ${\overline K}K$ in nanobarns.
 The upper numbers are for the integral up to 2~GeV$^2$, while the lower set includes the estimate of the contribution above 2~GeV$^2$.
$\;\dag$ For the $I=1$ channel we have included the contribution of the $a_2(1230)$ in italics. Unlike the states coupling to $\pi\pi$, this is not the result of an Amplitude Analysis, but is estimated in the pure helicity two Breit-Wigner approximation. The first error quoted for $a_2$ is that from $\gamma\gamma$ coupling quoted in ~\cite{pdg}; the second error is our estimate (from the determination of the \lq\lq correct'' $f_2(1270)$ contribution) of the uncertainty from the approximations made.} 
\end{center}
}
\end{table}

When the intermediate state is a resonance, its contribution is included in the sum of multiparticle modes to which it decays.  For instance, the tensor meson, the $f_2(1270)$ contributes
through its $\pi\pi$, ${\overline K}K$ and $4\pi$ channels. In the same narrow resonance approximation, a resonance of mass $M_R$ contributes to an integral of the $\gamma\gamma\to\pi\pi$ cross-section with helicity $\lambda$ as
\be
\sigma_\lambda(\gamma\gamma\to\pi\pi;s)\;=\;8\pi^2\,(2J+1)\frac{\Gamma_{\gamma\gamma}^{(\lambda)}}{M_R}\, {\rm BR}(R\to\pi\pi)\,\delta(s-M^2) \; ,
\ee
where BR is the branching ratio of the resonance decay, here to $\pi\pi$.
The difference of a factor of two between Eq.~(2) and Eq.~(3) comes about because of differences in the relation of the unpolarized cross-section to its helicity components.
Thus $\gamma\gamma\to\gamma\gamma$ cross-section~\cite{PV1} $\sigma\,=\,(\sigma_0 + \sigma_2)/2$ is normalized as in Eq.~(2), while the unpolarized $\gamma\gamma\to\pi\pi$ cross-section~\cite{lyd-mrp}
for each isospin $I$ is~{\footnote{The physical cross-sections for $\pi^+\pi^-$ and $\pi^-\pi^0$ are related to integrals of  the sums of the modulus squared of the helicity amplitudes of definite isospin and so involve interferences of isospin amplitudes. Only the sum of the $\pi^+\pi^-$ and $\pi^0\pi^0$ cross-sections (where the interference cancels) is simply related to $\sum_{I,\lambda=0,2} \,\sigma^I_\lambda$.}}  $\sigma^I\,=\,(\sigma^I_0+\sigma^I_2)$ as normalized in Eq.~(3). Since the partial cross-sections for $\pi\pi$ and ${\overline K}K$ in our amplitude analysis have been normalized according to Eq.~(3), we scale these results by a factor two to match the $\gamma\gamma$ cross-section in each helicity. Of course, the $f_2(1270)$, like the $f_0(500)$ and $f_0(980)$, is not well described by a narrow resonance approximation, so the contributions from our analysis of experimental data will not coincide with the pure helicity two approximation in ~\cite{PV1,PV2}.
Nevertheless, for the want of anything more definite, in the $I=1$ 3$\pi$ channel, where we have no amplitude analysis, we have estimated the contribution of the $a_2(1230)$ in the helicity two Breit-Wigner approximation and included this in Table~1, with a suitably expanded error.

At very low $\pi\pi$ masses the magnitude of the cross-sections ($I=0, 2$, $\lambda =0, 2$) is known to
be close (within ~30\%) to a one pion exchange Born model. Indeed  in this Born approximation the sumrule
can be integrated to infinite energy, and helicity-zero and two components do indeed contribute equally, as one can readily check
analytically --- see the Appendix.

Of course, the Born amplitude contains no strong interaction dynamics that dominates the contribution from hadronic intermediate states. To do better, one has to use the results of a partial wave separation of $\gamma\gamma$ scattering.
 This is the context for a recent coupled channel Amplitude Analysis~\cite{lyd-mrp} of the high statistics
results from Belle on $\gamma \gamma$ to two mesons $\pi\pi$~\cite{Belle-pm,Belle-nn} and $K\overline{K}$~\cite{Belle-KsKs} (and eventually $\pi^0 \eta$~\cite{Belle-pieta}).
Only where we have a partial wave separation can we know the result for the whole angular range), and even then the upper energy is far below infinity required to evaluate Eq.~({\ref{eq:PV1}).
Because of the energy range of the amplitude analysis,  we can only integrate from $\pi\pi$, or $\overline{K}K$ threshold to 2~GeV$^2$.

While our Amplitude Analysis has determined the $I=0, 2$ $\pi\pi$, and the $I=0, 1$ ${\overline K}K$  $S$ and $D_\lambda$ waves up to $s \simeq 2$ GeV$^2$, all the higher waves are approximated  by their one pion (or kaon) exchange amplitude, ${\cal B}_{J\ge 4}$.
%, plotted in two ways.
%The dashed lines are from these waves directly. To these we have to add the contribution from the higher waves in this same energy region, and the contribution from all waves above 1.44~GeV.
 Thus the amplitudes for each isospin (we suppress the label here) are
\be\label{eq:modBorn}
{\cal M_\lambda}(s,\theta,\phi) \;=\; {\cal S}_0(s)\, Y_{00}(\theta,\phi) \delta_{0\lambda} + {\cal D}_\lambda(s)\, Y_{2\lambda}(\theta,\phi) +  {\cal B}_{J\ge 4,\lambda}(s,\theta,\phi)
\ee
for $\,\sqrt{s} \le 1.44$~GeV. From these amplitudes we can deduce the helicity cross-section difference $\;\Delta\sigma(s) = \sigma_2(s) -\sigma_0(s)\;$ that appears in Eq.~(\ref{eq:PV1}). In Fig.~1 we show the integrands of the $I=0,2\,$ PV sumrules for each of these up to $s = 2$~GeV$^2$.
\begin{figure}[hp]
\vspace{-0.0cm}
\centering
\includegraphics[width=1.0\textwidth,height=0.70\textheight]{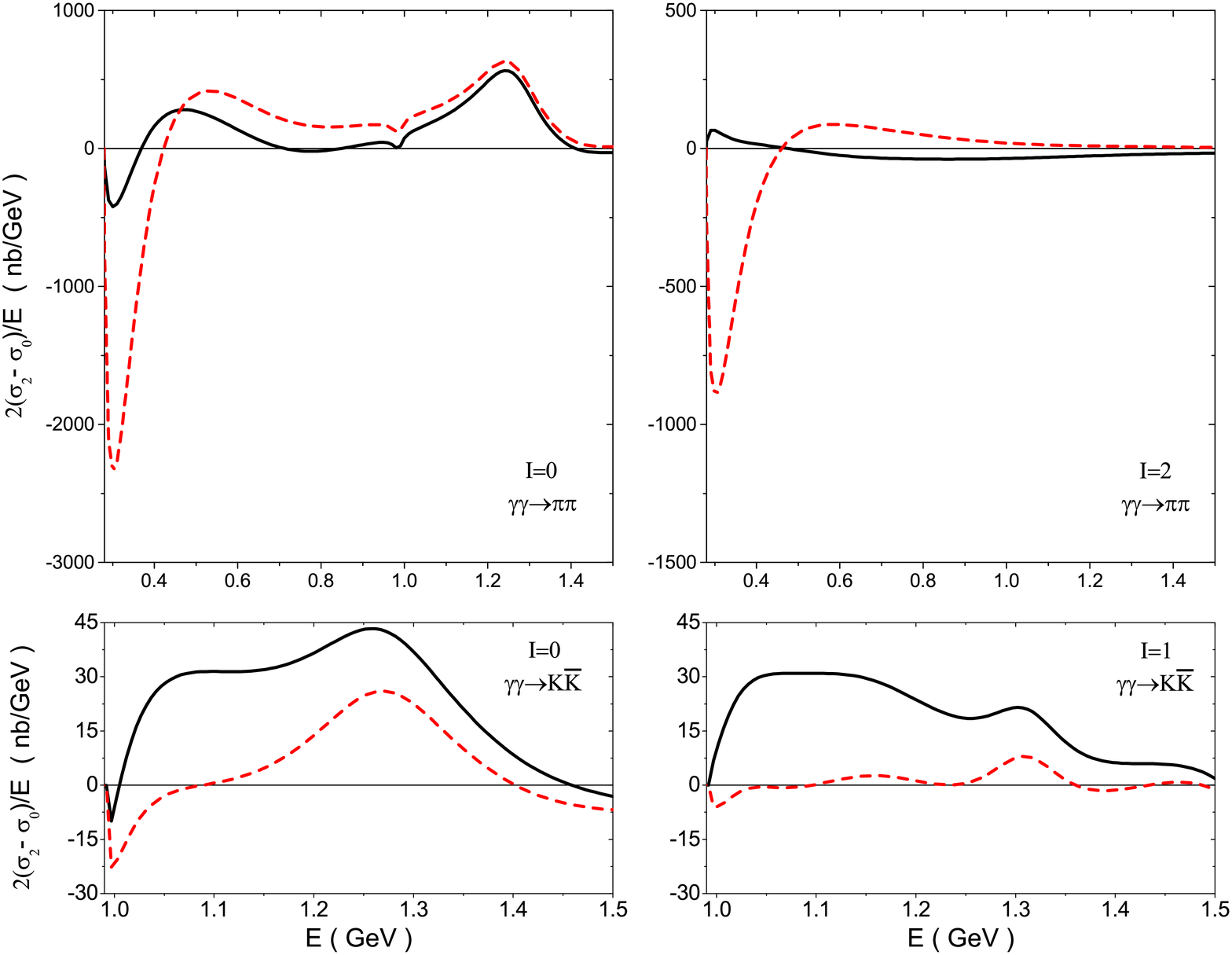}
\caption{\label{fig:PV} The contribution of isospin 0 and 2 cross sections to the integrands for the PV sumrule, Eq.~(1), for the difference
of helicity two and zero cross-sections, $\Delta\sigma$. Note the different ordinate scales for these plots. $\int ds\, (\sigma_2-\sigma_0)/s~ds$ of Eq.~(\ref{eq:PV1}) is, of course, the same as $\int dE\, 2(\sigma_2-\sigma_0)/E$, with $E$ the energy in the center of mass frame.   The dashed lines are for $\Delta\sigma$, the solid for ${\overline{\Delta\sigma}}$ of Eq.~(6), {\it i.e.} with the Born cross-section difference subtracted. In this latter case the PV sumrule essentially requires no contribution at higher energies from these channels. }
\end{figure}
The resulting contributions for $\pi\pi$ and ${\overline K}K$ intermediate states are shown in the top half of Table~1. That for the ${\overline K}K$ channel is generally much smaller than that for $\pi\pi$.
 While the Amplitude Analysis determines the $f_2(1270)$ is indeed dominated by its helicity-two component, it does have a helicity-zero component of ~$(8.6 \pm 1.7)\%$ and a substantial $S$-wave cross-section in the same mass region.
%Consequently, the region from 1150 to 1300~MeV makes a contribution half (??) that implied by a pure helicity-two $f_2$ in the narrow resonance approximation.

While the spin zero and two waves are distinctly different from the Born approximation, reflecting important direct channel dynamics, we know that an infinity of higher waves must be very close to the Born amplitude for $\pi\pi$ production reflecting the closeness of the $t$ and $u$-channel pion poles to the physical region.
Thus for instance at $\sqrt s =2$~GeV, the pion poles are at $\cos\theta = \pm1.01$, only just outside the physical region. The amplitudes ${\cal M}$ reflect this, Eq.~(\ref{eq:modBorn}).  In contrast for ${\overline K}K$ production, the kaon poles are much further away, being at $\cos\theta = \pm 1.15$, again at $\sqrt s =2$~GeV. Thus the Born approximation is there poorer.

Nevertheless, these considerations provide the motivation for our estimate of the higher energy contribution to the PV sumrule.
These can be calculated by using the Born amplitude as a reasonable
approximation for $s > 2$~GeV$^2$. For the $\pi\pi$ channels, studies with different high energy behavior suggest that this is accurate to about 10\%, while for ${\overline K}K$ to 25\%. However, the total contribution to the PV sumrule from the kaon channel is much smaller than that of $\pi\pi$, and so its larger uncertainty matters less.
In adding the high energy contribution, we can profit from the fact that the PV sumrule is exactly satisfied by the Born amplitude. Consequently 
%with $\Delta\sigma(s) \equiv (\sigma_2(s)-\sigma_0(s))$
\be
\int_{s_{th}}^S\,ds\,\frac{\Delta\sigma_{Born}(s)}{s}\;=\;-\,\int_S^{\infty}\,ds\,\frac{\Delta\sigma_{Born}(s)}{s}\quad.
\ee
Thus the total integrand for the PV sumrule can be expressed wholly as an integral from $s=s_{th}$ to $S =$ 2~GeV$^2$
of just the $S$ and $D_\lambda$ partial waves with
\be
{\overline \Delta\sigma}^I(s)\;=\; \sigma^I_{D2}(s) - \sigma^I_S(s) - \sigma^I_{D0}(s) - \left[\sigma^I_{D2}(s) - \sigma^I_S(s) - \sigma^I_{D0}(s)\right]_{Born}\quad \\[1.5mm].
\ee
The components of this integrand are also shown  in Fig.~1 as the solid lines. The result of this integral is also given in Table~1. Our recalculation of $I=0$ $\gamma\gamma\to\eta$, $\eta'$, $\pi\pi$, and $K\overline{K}$ is thus  (-166$\pm$46)~nb, with the systematic error of 28.0\%, and $(-44\pm 6)$~nb with a 14\% systematic uncertainty for $I=2$. Clearly these are not zero.
Consequently, there must be additional intermediate states that can
make a substantial contribution in the few GeV region.

Before considering such contributions,
it is helpful for this discussion to define the  contributions of two particle intermediate states to $\gamma\gamma\to\gamma\gamma$ in terms of differential cross-sections, where $z\,=\,\cos\theta^*$, with $\theta^*$ the c.m. scattering angle. Then for each isospin $I$, we have
\be
\label{eq:SumX}
\Sigma^I(s_1,s_2,Z, {\rm channel})\;=\;
%\frac{1}{2}\,
\int_{s_1}^{s_2}\,\frac{ds}{s}\,\int_{-Z}^Z\, dz\, \left(\frac{d}{dz}\sigma^I_2\ +\frac{d}{dz}\sigma^I_0 \right)\quad.
\ee
This is a quantity that can be deduced from measurements with unpolarized photons. Closer to the PV sumrule is the difference, $\Delta$, rather than this sum, $\Sigma$. This we define by
\be
\label{eq:PV2}
\Delta^I(s_1,s_2,Z, {\rm channel})\;=\;
%\frac{1}{2}\,
\int_{s_1}^{s_2}\,\frac{ds}{s}\,\int_{-Z}^Z\, dz\, \left(\frac{d}{dz}\sigma^I_2\ -\frac{d}{dz}\sigma^I_0 \right)\quad.
\ee
 The multiparticle  contributions to these can only be deduced after an Amplitude Analysis. The sumrule of Eq.~(\ref{eq:PV1}), of course, requires $s_1 = 4m_\pi^{\,2}$, $s_2 = \infty$, $Z = 1$.
We usefully define the ratio, ${\cal R}$,
\be
\label{eq:ratio}
{\cal R} (s_1, s_2; {\rm channel})\;=\; \frac{\Delta(s_1, s_2, Z=1; {\rm channel})}{\Sigma(s_1, s_2, Z_{\rm exp}; {\rm channel})} \quad\\[1.5mm].
\ee
This provides a scaling factor with which to multiply the experimental cross-sections, to estimate, their contribution to the PV sumrule.
%\newpage
\section{Contributions to the Sumrules: $4\pi$, {\it etc.}}

Published data~\cite{compilation} allow the contributions to the integral, $\Sigma$, of the sum of cross-sections, Eq.~(\ref{eq:SumX}),
for $\gamma\gamma\to$ multi-meson processes~\cite{data1}-\cite{data8}, 4$\pi$, $\pi\pi K \overline{K}$, $\cdots$ to be computed. These are listed in Table~\ref{tab:sig}.
That these are large means such intermediate states
will contribute significantly to light-by-light scattering and probably to the PV sumrules too. The ratio ${\cal R}$ of Eq.~(\ref{eq:ratio})
 provides a scaling factor, with which to multiply the experimental cross-sections, to estimate their contribution to the PV sumrule.
% It is worth repeating that this requires an integral over the full angular range with $Z=1$.
We estimate this scaling factor in two ways.
\begin{itemize}
\item[(i)] From our Amplitude Analysis we know the ratio, ${\cal R}_{\rm AMP}$ in each charged channel, and also separated by isospin, but only for $s$ from $\pi\pi$ threshold to 2~GeV$^2$,
\item[(ii)] From the $\pi\pi$ Born amplitude integrated over the defined range of energies $s_1 \le s \le s_2$. An example of the calculation involved is set out in the Appendix.
\end{itemize}
Assessment (i) typically gives ${\cal R}\simeq 0.65$ from $s=1$ to 2~GeV$^2$. However, this is in the region where the helicity-zero component is largest. From 2 to 4~GeV$^2$, which we need to assess the contribution of the multi-pion data, this ratio goes above one, as we now discuss.
Assessment (ii) uses the Born approximation. Then the sum and difference of the differential helicity cross-sections integrated up to $\cos\theta =Z$ and from threshold $s_{th}$  to energy squared $S$  with $X^2 = 1- s_{th}/S$ are respectively
\bea
\label{eq:SigBorn}
\Sigma_{Born}(s_{th},S,Z)& =& \frac{e^4}{2s_{th}}\, \Big\{\frac{1}{12Z^6} \left [ 5 -12 Z^2 +9 Z^4 -(3X^2 -X^6) Z^6\right]\, \ln\left(\frac{1+XZ}{1-XZ}\right)\no\\
&&+ \frac{X}{6Z^5}\left[ -9Z^4 +12 Z^2 - 5 \right] +\frac{X^3}{18Z^3}\left[ 6 Z^4 +12 Z^2 -5\right] - \frac{X^5}{6Z}\Big\}\quad ,\\[3mm]
\label{eq:DiffBorn}\Delta_{Born}(s_{th},S,Z)& =& \frac{e^4}{2s_{th}}\, \Big\{\frac{1}{4Z^4} [1-X^2Z^2]\,\left[ 2Z^2 -1 -X^2Z^2\right] \ln\left(\frac{1+XZ}{1-XZ}\right)\no\\
&&- \frac{X}{2Z^3} (2Z^2 -1) + \frac{X^3}{6Z} (2Z^2+1)\Big\}\quad .
\eea

\begin{table}[h]
\vspace{-0.0cm}
{\footnotesize
\begin{center}
\begin{tabular}  {||l|c||c|c|c|c||}
\hline
  Channel & Publication & $E_1$ (GeV) & $E_2$ (GeV)   &   $\Sigma$ (nb) & ${\cal R}(Born)$  \rbox \\[0.5mm]\hline\hline
$\pi^+\pi^-$ ($Z=0.6$) & \cite{data1}& 2.4 & 4.1 & 0.44 $\pm$ 0.01 & 1.61 \rbox\\[0.5mm]\hline
$K^+K^-$ ($Z=0.6$) & \cite{data1} & 2.4 & 4.1 & 0.39 $\pm$ 0.01& 1.29 \rbox\\[0.5mm]\hline
 $\pi^0\pi^0$ ($Z=0.8$) & \cite{data2} & 1.44 & 3.3 & 8.8 $\pm$ 0.2 & 1.18 \rbox\\[0.5mm]\hline
$\pi^0\pi^0\pi^0$ & \cite{data3} & 1.525 & 2.425 & 5.8 $\pm$ 0.8 &1.55 \rbox \\[0.5mm] \hline
$\pi^+\pi^-\pi^0$ (non-res.) & \cite{data4} & 0.8 & 2.1 & 23.0 $\pm$ 1.3 & 1.39\rbox \\[0.5mm] \hline
$K_s K^{\pm}\pi^{\mp}$ & \cite{data5} & 1.4 & 4.2 & 9.7 $\pm$ 1.6 &  \rbox \\[0.5mm] \hline
$\pi^+\pi^-\pi^+\pi^-$ & \cite{data6} & 1.1 & 2.5& 215 $\pm$ 11 $\pm$ 21        & 1.49       \rbox \\[0.5mm] \hline
$\pi^+\pi^-\pi^+\pi^-$ & \cite{data7} & 1.0 & 3.2 & 153 $\pm$ 5 $\pm$ 39   & 1.48  \rbox \\[0.5mm] \hline
$\pi^+\pi^-\pi^0\pi^0$ & \cite{data8} & 0.8 & 3.4 & 103 $\pm$ 4 $\pm$ 14 &1.42 \rbox \\[0.5mm] \hline

\end{tabular}
\caption{\label{tab:sig} Integral of channels specified to Eq.~(\ref{eq:SumX}) from $s_1=E_1^{\,2}$ to $s_2=E_2^{\,2}$ as listed. Note these cross-sections are not separated
for either isospin or helicity.   They are the sum of all contributions, except for the $3\pi$ denoted by \lq non-res.' from which the experimental analysis has removed the $a_2(1320)$ contribution.
The factor ${\cal R}$ defined from Eqs.~(\ref{eq:ratio}-\ref{eq:DiffBorn}) is an \lq\lq estimate'' of the factor, by which the listed cross-sections $\Sigma$ need to be scaled to give the contribution of each channel and energy region to the PV sumrule --- see text for the discussion.}
\end{center}
}
\end{table}

From Eqs.~(\ref{eq:SigBorn}, \ref{eq:DiffBorn}) we can then deduce the ratio ${\cal R}$ defined in Eq.~(\ref{eq:ratio}) from the Born amplitude listed in Table~2. We see that this enhances the expected contribution to the PV sumrule. This may appear strange given that the difference of the helicity-two and helicity-zero cross-sections of Eq.~(\ref{eq:PV2}) is surely less than the sum of these cross-sections. The reason this is not the case is because as the energy increases %
% (and not all the examples in Table~2 have the lower energy  $E_1 > 0,8$~GeV)
the sumrule for $\Delta$ is  dominated by the helicity-two contribution and this has the biggest difference between $Z\sim 0.6$ and $Z=1$, {\it cf.} Eq.~(\ref{eq:ratio}). Helicity-zero contributes most to the $S$-wave and this is only large at low energies (remember the integral in Eq.~(\ref{eq:PV1}) has a factor $1/s$ in the measure in addition to the natural decrease in the cross-section at higher energies). This is seen in the negative contributions in Fig.~1. While both the integrals defined for the cross-section sum and  difference by Eqs.~(\ref{eq:SumX}, \ref{eq:PV2}), as with the PV sumrule, Eq.(\ref{eq:PV1}), are dominated by contributions from low energies, their convergence is not so very fast. Using the Born amplitude as a guide, Eq.~(\ref{eq:SigBorn}), $\Sigma(4m_\pi^{\,2},S,Z=1)$ reaches 90\% of its asymptotic value already by $\sqrt{S}$ of 1.25~GeV, and achieves 96\% by 2~GeV, but 98\% by 3~GeV. The difference, Eq.~(\ref{eq:PV2}), or rather the normalized ratio ${\cal R}$ is 10\% at 1.25~GeV, falling to 4\% at 2~GeV and  below 2\% at 3~GeV, on its way to zero asymptotically. To repeat. this is critically dependent on covering the whole angular range to $Z=1$.  

The region of 2--3~GeV, above the range of our Amplitude Analysis, being so important makes the large multi-meson cross-sections seen in Table~2 matter for the PV sumrule, with its required delicate cancellation.
The contribution to the sumrule from these multi-pion channels can be crudely estimated by taking the measured cross-sections, $\Sigma$, multiplying them by the ratio ${\cal R}$ that we have listed in Table~2, and scaling by normalization factors and guesses of the isospin decomposition, {\it i.e.} multiplying by a crude factor of $\sim 1$ for $I=0$ and $\sim 0.5$ for $I=2$.  This would suggest that these would readily contribute
the 150--200 nb in the $I=0$ channel and 50 nb in the $I=2$ mode. When added to our results in Table~1, these would make the integral in Eq.~(1) consistent with zero, as expected.

 Of course, the Born estimates know nothing of the direct channel dynamics that control $\gamma\gamma\to \rho\rho, \, \omega\pi,\, \omega\rho,\, \omega\omega,\, \cdots$.  As remarked earlier the Born approximation gives the right order of magnitude for the $\gamma\gamma\to\pi^+\pi^-$ cross-section in the low energy region, even though what is
observed experimentally is modified by substantial corrections from final state interactions, particularly in the $I=0$ channel. This rough agreement is because the pion poles at $t = u =m_\pi^{\,2}$ are very close to the $s$-channel physical region even at low energies. In contrast the kaon poles at $\, t = u = m_K^{\,2}$ are far from the physical region for $4m_K^{\,2} < s < 2$~GeV$^2$. Consequently, other $t$ and $u$-channel exchanges, like the $K^*(890)$ and the $\kappa/K^*_0(650)$ are just as important. %Indeed, the physical cross-section is an order of magnitude bigger than the simple Born amplitude being controlled by $s$-channel dynamics with the closeness of the $f_0$ and $a_0(980)$, and the near approach of the $a_2(1320)$.
 This situation is even more so for the $\rho^+\rho^-$ production, where estimates from the one pion exchange Born cross-section are more than order of magnitude below the observed cross-sections, since at threshold when $s=4m_\rho^{\,2}$ and $t=u=-m_{\rho}^{\,2}$ is very far from $t=u=m_\pi^{\,2}$. Indeed, long ago Achasov {\it et al.}~\cite{achasov-4q} proposed that the large $\rho\rho$ cross-section was dominated by the production of several wide tetraquark resonances. While this cannot be checked without a partial wave analysis, the proposal indicates the key role of direct channel dynamics in this crucial mass region for the data to satisfy the PV sumrule.

\section{Conclusion}
In this paper we set out the contributions to the PV sumrule for light-by-light scattering. Single, near {\it stable}, pseudoscalar mesons plus $\pi\pi, K\overline{K}$ intermediate states up to 1.44~GeV in $\gamma\gamma$ cm energy
contribute  (-166$\pm$46)~nb, with a systematic error of 28\% in the isoscalar channel and (-44$\pm$6)~nb, with a systematic error of 14\%, in the isotensor mode. These calculations are made possible by the recent Amplitude Analysis~\cite{lyd-mrp} in this energy region of the high statistics $\pi^+\pi^-, \pi^0\pi^0, K^0_SK^0_S$ data from Belle. We show that narrow resonance estimates from the tensor mesons are not a good approximation. Though the accurately determined contributions do not saturate the Pascalutsa-Vanderhaeghen sumrule for isospin zero or two, we find that it is most likely the four pion intermediate state that provides sufficient contribution below ~2.5~GeV to give the expected zero result. 

While there are data on the cross-sections for $\pi^+\pi^-\pi^+\pi^-$ and $\pi^+\pi^-\pi^0\pi^0$ production in the required energy region, there is insufficient information to do more than \lq\lq guestimate'' the isospin and helicity decomposition of these integrated data. All other contributions are small, at the few nanobarn level. Only four pion production delivers the missing  150--200~nb in the $I=0$ channel and 50~nb with $I=2$. Speculations of wide tetraquark states would render this quite natural~\cite{achasov-4q}.

Since  these sumrules play a key role in constraining the contribution of light-by-light scattering to $(g-2)$ of the muon, we urge experiments at $e^+e^-$ colliders, such as BESIII@BEPC~II and Belle@KEKB, to consider investing in detailed studies of 4$\pi$ production from untagged two photon data. Differential cross-sections for $\rho^+\rho^-$ and $\rho^0\rho^0$ production from threshold to ~2.5 GeV, even without helicity separation, would be a most useful guide in checking the expectations in this paper, and so testing the validity and utility of the simplest Pascalutsa-Vanderhaeghen sumrule.

\vspace{1cm}
\newpage

\noindent{\bf Acknowledgment}

\noindent We are grateful to Zaarah Mohamed, a summer student on the DOE SULI scheme at Jefferson Lab in 2014, who made various estimates of the high energy contribution from two pseudoscalar mesons to the sumrules we consider  here. These calculations were helpful in leading us to understand that only the multi-pion channels, in particular $4\pi$, had sufficient strength to saturate the light-by-light sumrules. We also thank Marc Vanderhaeghen for useful discussions. This work is supported in part by the
Deutsche Forschungsgemeinschaft (Grant No. SFB/TR 110, “Symmetries and the Emergence of Structure in
QCD”). We acknowledge support from Indiana University College of Arts and Sciences, and from the
U.S. Department of Energy, Office of Science, Office of Nuclear Physics under Contract No. DE-AC05-06OR23177,
which funds Jefferson Lab research.
%\newpage
\vspace{1cm}
\appendix
\section{Appendix}
From the one pseudoscalar meson exchange Born amplitude, we can estimate from the known cross-section for the sum of helicities integrated over a limited angular range, $\Sigma(s_1,s_2,Z)$, Eq.~(\ref{eq:SumX}),
what
 the helicity difference integrated over the full angular range that enters the Pascalutsa-Vanderhaeghen sumrule~\ref{eq:PV1} is. Here we set out part of the calculation. First recall that the helicity anplitudes, ${\cal M}_{\lambda1\lambda2}$ with $\lambda_1\lambda_2\,=\,++$ or $+-\,$ for $\gamma\gamma\to\pi\pi$ in the center-of-mass frame are related to the differential cross-sections by
\be
\frac {d\sigma_{\lambda1\lambda2}}{d\cos \theta^*}\;=\;\frac{\beta}{128\pi^{\,2}\ s} \,\int_0^{2\pi}\,d\phi\, |{\cal M}_{\lambda1\lambda2}(s,\cos\theta^*,\phi)|^2 \quad ,
\ee
where for a meson of mass $m$, $\beta\,=\,\sqrt{1 =4m^2/s}$. Note that in Eqs.~(3,4) the total helicity $\lambda =\lambda_1-\lambda_2$. In the Born approximation, these helicity amplitudes are given by
\bea
{\cal M}_{+-}(s, \theta^*, \phi)&=&e^2\,\sqrt{16\pi}\; \frac {\beta^2\ \sin^2 \theta^*}{1 - \beta^2 \cos^2 \theta^*}\, {\exp 2i\phi}\nn\\
{\cal M}_{++}(s, \theta^*, \phi)&=&e^2\, \sqrt{16\pi}\; \frac {{1 - \beta^2}}{1 - \beta^2 \cos^2 \theta^*}\quad ,
\eea
with $e$ the charge of the pion in units in which $\hbar=c=1$.
Then  on integrating the square of their moduli over $\phi$, we have writing $\cos\theta^* = z$
\be
\frac{d}{dz}\sigma_{+-}\,-\,\frac{d}{dz}\sigma_{++}\;=\; e^4\,\frac {\beta}{4 s}\;\left[1\;-\;\frac{2(1-\beta^2)}{1\ -\ \beta^2\ z^2}\right]\quad .
\ee
Integrating over $z$ up to value $Z$ gives
\be
\sigma_{+-}(s,Z)\;-\;\sigma_{++}(s,Z)\;=\;\frac{e^4}{4}\,\frac{\beta}{s}\,\left[ Z - \frac{(1-\beta^2)}{\beta}\, \ln\left(\frac{1+\beta Z}{1-\beta Z}\right)\right]\quad.
\ee
Changing integration variable from $s$ to $x \equiv\beta$, and noting $ds\,=\,x\,dx\, s^2/(2 m^2)$, we have integrating from $x=0$ to $x=X=\sqrt{1-4m^2/S}$
\bea
\Delta_{Born}(s_{th},S,Z)&=&\int_{4 m^2}^S \; \frac{ds}{s}\;\left[ \sigma_{+-}(s) - \sigma_{++}(s) \right]\\
&=&\frac{e^4}{8 m^2} \Big\{\frac{1}{4Z^4} [1-X^2Z^2]\,\left[ 2Z^2 -1 -X^2Z^2\right] \ln\left(\frac{1+XZ}{1-XZ}\right)\no\\
&&\hspace{3.6cm}- \frac{X}{2Z^3} (2Z^2 -1) + \frac{X^3}{6Z} (2Z^2+1)\Big\}\quad .
\eea

\noindent Similar integration gives the equation for $\Sigma_{Born}(s_{th},S,Z)$ shown in Eq.~({\ref{eq:SigBorn}). These are used in the establishing the estimates for the ratio ${\cal R}$, Eq.~(9), in the final column of Table~2.

Note that when $Z=1$, Eq.~(17) becomes
\be
\Delta_{Born}(s_{th},S,Z=1)\;=\;\frac{e^4}{32 m^2}\,(1-X^2)\left[(1-X^2)\,\log\left(\frac{1 + X}{1 - X}\right) - 2 X\right] \quad .
\ee
Of course, when $S \to \infty$ ({\it i.e} $X \to 1$), $\Delta_{Born} \to 0$, but only if integrating over the full angular range.
%\newpage
%\appendix

\vspace{1cm}
%%%%%%%%%%%%%%%%%%%%%%%%%%%%%%%%%%%%%%%%%%%%%%%%%%%%%%%%%%%%%%%%%%%%%%%%%%%%%%%%%%%%%%%%%%%%%%%%%%%
%\newpage

%%%%%%%%%%%%%%%%%%%%%%%%%%%%%%%%%%%%%%%%%%%%%%%%%%%%%%%%%%%%%%%%%%%%%%%%%%%%%%%%%%%%%%%%%%%%%%%%%%%

\end{document}